\documentstyle[aps,preprint,pra,epsfig]{revtex}

\begin{document}

\draft

\title{On Defect-Mediated Transitions in Bosonic Planar Lattices}

\author{A. Trombettoni$^{1}$, A. Smerzi$^{2,3}$, and P. Sodano$^{4}$}
\address{$^1$ Istituto Nazionale per la Fisica della Materia and
Dipartimento di Fisica, Universita' di Parma,
parco Area delle Scienze 7A, I-43100 Parma, Italy\\
$^2$ Istituto Nazionale per la Fisica della Materia BEC-CRS and
Dipartimento di Fisica, Universita' di Trento, I-38050 Povo, Italy\\
$^3$ Theoretical Division, Los Alamos
National Laboratory, Los Alamos, NM 87545, USA\\
$^4$ Dipartimento di Fisica and Sezione I.N.F.N., Universit\`a di 
Perugia, Via A. Pascoli, I-06123 Perugia, Italy\\
}
\date{\today}
\maketitle

\begin{abstract}
We discuss the finite-temperature properties of Bose-Einstein condensates 
loaded on a $2D$ optical lattice. In an experimentally attainable range 
of parameters the system is described by the $XY$ model, which undergoes 
a Berezinskii-Kosterlitz-Thouless (BKT) transition driven by the 
vortex pair unbinding. 
The interference pattern of the expanding condensates provides 
the experimental signature of the BKT transition: 
near the critical temperature, the $\vec{k}=0$ component 
of the momentum distribution sharply decreases.
\end{abstract}


\newpage

Bose-Einstein condensates (BECs) are nowadays routinely 
stored in $1D$ 
\cite{anderson98,cataliotti01,morsch01,hensinger01,eiermann03},  
$2D$ \cite{greiner01} and 
$3D$ \cite{greiner02} optical lattices. 
In this paper we review and further discuss  
the finite-temperature properties of Bose-Einstein condensates 
loaded on a $2D$ optical lattice \cite{cm}: there we showed 
that at a critical temperature 
lower than the temperature $T_{BEC}$ at which condensation 
in each well occurs, $2D$ lattices of BECs may undergo a
phase transition to a superfluid regime 
where the phases of the single-well condensates are coherently aligned
allowing for the observation of a 
Berezinskii-Kosterlitz-Thouless (BKT) transition. 

The BKT transition is the paradigmatic example of a 
defect-mediated transition \cite{nelson83} 
which is exhibited by the two-dimensional 
$XY$ model \cite{minnaghen87,kadanoff00}, 
describing $2$-components spins on a two-dimensional lattice. 
In the low-temperature phase, characterized by the presence of 
bound vortex-antivortex pairs, the spatial correlations exhibit a power-law 
decay; above a critical temperature $T_{BKT}$, the decay is exponential 
and there is a proliferation of free vortices. 
The BKT transition has been observed in superconducting Josephson arrays 
\cite{resnick81} and its predictions are well verified in 
the measurements of the superfluid density 
in $^4He$ films \cite{bishop78}. In finite magnetic systems 
with planar symmetry \cite{bramwell93}, the BKT transition point 
is signaled by the drop of a suitably defined
magnetization \cite{berezinskii73}. We shall discuss in the sequel 
the analogous of this magnetization for atomic systems.  

For $2D$ optical lattices, when the polarization vectors of the two 
standing wave laser fields are orthogonal, the resulting periodic 
potential for the atoms is
$V_{opt}=V_0 [ \sin^2{(kx)} + \sin^2{(ky)} ]$
where $k=2 \pi/\lambda$ is the wavevector of the laser beams. 
The potential maximum of a single standing wave 
$V_0=sE_R$ may be controlled by changing the intensity of the optical 
lattice beams, and it is conveniently measured in units of the recoil energy
$E_R=\hbar^2 k^2/2m$ ($m$ is the atomic mass), 
while, typically, $s$ can vary from $0$ up to $30$. Gravity 
is assumed to act along the $z$-axis. 
Usually, superimposed to the optical potential 
there is an harmonic magnetic potential  
$V_{m}=(m/2) [ \omega_r^2 (x^2+y^2) + \omega_z^2 z^2 ]$,
where $\omega_z$ ($\omega_r$) is the axial (radial) trapping frequency. 
The minima of the lattice
are located at the points $\vec{j}=(j_1,j_2) \cdot \frac{\lambda}{2}$ 
with $j_1$ and $j_2$ integers, and the potential around the minima is
$V_{opt} \approx 
\frac{m}{2} \tilde{\omega}_r^2 [ (x-\lambda j_1/2)^2 +  
(y-\lambda j_2/2)^2 ]$ with $\tilde{\omega}_r=\sqrt{2 V_0 k^2 / m}$. 
When $\tilde{\omega}_r \gg \omega_r,\omega_z$, 
the system realizes a square array of tubes, i.e. an array of harmonic traps 
elongated along the $z$-axis \cite{greiner01}. 
In the following we analyze only 
the situation in which the axial degrees of freedom are frozen out, 
which is realizable if $\omega_z$ is sufficiently large. We also
assume that the harmonic oscillator length 
of the magnetic potential $\sqrt{\hbar/m \omega_r}$ is 
enough larger than the size $L$ (in lattice units) of the optical 
lattice in order to reduce density inhomogeneity effects and to allow for 
a safe control of finite size effects.

When all the relevant energy scales 
are small compared to the excitation 
energies one can expand the field operator \cite{jaksch98} as 
$\hat{\Psi}(\vec{r},t)= \sum_{\vec{j}} \hat{\psi}_{\vec{j}}(t) 
\Phi_{\vec{j}}(\vec{r})$ with $\Phi_{\vec{j}}(\vec{r})$ 
the Wannier wavefunction localized in the $\vec{j}$-th well 
(normalized to $1$) and $\hat{N}_{\vec{j}}=\hat{\psi}^{\dag}_{\vec{j}} 
\hat{\psi}_{\vec{j}}$ the bosonic number operator.
Substituting the expansion in the full quantum Hamiltonian 
describing the bosonic system, leads to 
the Bose-Hubbard model (BHM) \cite{fisher89,jaksch98} 
\begin{equation} 
\hat{H}=-K \sum_{<\vec{i}, \vec{j}>} 
(\hat{\psi}^{\dag}_{\vec{i}} \hat{\psi}_{\vec{j}}+ h.c.)
+ \sum_{\vec{j}} \Bigg[ {U \over 2} \hat{N}_{\vec{j}} 
(\hat{N}_{\vec{j}}-1) \Bigg]
\label{B-H}
\end{equation}
where $\sum_{<\vec{i}, \vec{j}>}$ denotes a sum over 
nearest neighbours, $ U = (4 \pi \hbar^2 a / m) \int d\vec{r} \, 
\Phi_{\vec{j}}^4$ ($a$ is the $s$-wave scattering length) and 
$K \simeq - \int d\vec{r} \, \big[ \frac{\hbar^2}{2m} 
\vec{\nabla} \Phi_{\vec{i}} \cdot \vec{\nabla} \Phi_{{\vec{j}}} + 
\Phi_{\vec{i}}  V_{ext} \Phi_{\vec{j}} \big]$. 
 
Upon defining $J \approx 2 K N_0$ (where $N_0$ is the 
average number of atoms per site),   
when $N_0 \gg 1 $ and $J/N_0^2 \ll U$, 
the BHM reduces to 
$\hat{H}=H_{XY}- {U \over 2}\sum_{\vec{j}} 
\frac{\partial^2}{\partial \theta_{\vec{j}}^2}$, 
which describes the so-called quantum phase model 
(see e.g. \cite{simanek94,fazio01}): 
$\theta_{\vec{j}}$ is the phase of the $j$-th condensate and 
\begin{equation}
H_{XY}=-J \sum_{<\vec{i},\vec{j}>} \cos{(\theta_{\vec{i}}-
\theta_{\vec{j}})}
\label{X-Y}
\end{equation}
stands for the Hamiltonian of the classical $XY$ model. 
When $J \gg U$ and at temperatures $T \gg U/k_B$, 
the quantum phase Hamiltonian may be 
well approximated by Eq.(\ref{X-Y}); 
therefore, the pertinent partition function 
describing the thermodynamic behaviour 
of the BECs stored in an optical lattice can be computed 
in these limits with the classical $XY$ model.

We remark that the BHM - for $U=0$ -  
describes harmonic oscillators and 
thus cannot sustain any BKT transition. However, 
in the limits above specified ($N_0 \gg 1 $ and $J/N_0^2 \ll U \ll J$) 
the Bose-Hubbard Hamiltonian reduces in the $XY$ model 
(\ref{X-Y}), which displays the BKT transition. We observe that in 
the chosen range of parameters, the condition 
$U N_0^2/J=U N_0/2K \gg 1$, satisfies the 
finite-temperature stability criterion 
recently derived in \cite{tsuchiya03}.  
In superconducting networks 
$N_0$ is the average number of excess Cooper pairs and 
thus, the requirements $N_0 \gg 1 $ and $J/N_0^2 \ll U$ are 
always satisfied; at variance, in bosonic lattices 
$N_0$ varies usually between $\sim 1$ and $\sim 10^3$ and the validity 
of the mapping in the quantum phase Hamiltonian is not always guaranteed.      
  
A simple estimate of the coefficients $J$ and 
$U$ may be obtained by 
approximating the Wannier functions with gaussians, 
whose widths are determined variationally \cite{cm}. 
For a $2D$ lattice with $V_0$ between $20$ and $25E_R$ (and 
$N_0 \approx 170$ as in \cite{greiner01}), 
the conditions $J \gg U \gg J/N_0^2$ are rather well 
satisfied and the BKT critical temperature, $T_{BKT} \sim J/k_B$, 
is between $10$ and $30nK$. 

We observe that, 
using a coarse-graining approach to determine the finite-temperature  
phase-boundary line of the Bose-Hubbard model (\ref{B-H}) 
\cite{kampf93}, 
one gets - in $2D$ for $N_0 \gg 1$ and $J / U \gg 1 $ - 
a critical temperature $T_{BKT} \approx J / k_B$. 
Indeed, within a coarse-graining approach, the 
phase-boundary line of the Bose-Hubbard model at a temperature 
$T$ is given by \cite{kampf93}
\begin{equation}
1=K d \int_{0}^{\beta} d \tau G(\tau)
\label{PBL}
\end{equation}
where $d$ is the lattice dimension and 
$\beta=1/k_B T$. In Eq.(\ref{PBL}) $G(\tau)$ is the correlation function, 
which, for $N_0 \gg 1$, is given by
\begin{equation}
G(\tau) \approx N_0 \frac{\vartheta_3 \bigg( 
\pi(N_0+\tau/\beta),q \bigg)}{\vartheta_3 \bigg( 
\pi N_0,q \bigg)} e^{-\tau U(1-\tau/\beta)/2}
\label{correlatore}
\end{equation}
with $\vartheta_3(z,q)=1+2 \sum_{n=1}^{\infty} \cos{(2nz)} 
\cdot q^{n^2}$. Since $d=2$, and $J=2KN_0$, 
from Eq.(\ref{PBL}) it follows $1=J G(\omega=0)/N_0$ where 
$G(\omega=0)=\int_0^\beta d \tau G(\tau)$; one has then  
\begin{equation}
\frac{G(\omega=0)}{N_0}=
\frac{\sum_{m} \exp{(-\beta U \phi_n^2 / 2)} \cdot 
\frac{1-\exp{(-\beta U (\phi_n + 1 / 2)}}{U(\phi_m+1/2)}} 
{\sum_m \exp{(-\beta U \phi_n^2 / 2)} }:
\label{trasf_corr}
\end{equation}
where $\phi_m=N_0-m$. Eq.(\ref{trasf_corr}) yields, for $U \ll k_B T$,  
$G(\omega=0) / N_0 \approx \beta$, thus implying 
$\beta J \approx 1$, i.e. $k_B T_{BKT} \approx J$, 
in fair agreement with $XY$ estimates.

We notice that the $XY$ Hamiltonian, 
in the limits $N_0 \gg 1$ and $J \gg U \gg J/N_0^2$, can be retrieved  
extending the Gross-Pitaevskii (GP) Hamiltonian to 
finite temperature $T \gg U/k_B$. 
In fact, replacing the tight-binding ansatz 
$\Psi(\vec{r},t)= \sum_{\vec{j}} \psi_{\vec{j}}(t) 
\Phi_{\vec{j}}(\vec{r})$ (where $\psi_{\vec{j}}$ are classical fields 
and not operators) in the GP Hamiltonian 
one gets the lattice Hamiltonian 
$H=-K \sum_{<\vec{i}, \vec{j}>} 
(\psi^{\ast}_{\vec{i}} \psi_{\vec{j}}+ c.c.)
+ (U/2) \sum_{\vec{j}} N_{\vec{j}} (N_{\vec{j}}-1)$, 
which is the classical version of the Bose-Hubbard model 
($N_{\vec{j}} \equiv \mid \psi_{\vec{j}}\mid^2$). 
Writing $N_{\vec{j}}=N_0 + \delta N_{\vec{j}}$, one may 
neglect 
the quadratic terms [i.e. $(\delta N_{\vec{j}})^2$] 
in the hopping part of the Hamiltonian 
$-K \sum_{<\vec{i}, \vec{j}>} 
(\psi^{\ast}_{\vec{i}} \psi_{\vec{j}}+ c.c.)$, which then 
reduces to $H_{XY}$ (\ref{X-Y}). 

Accurate Monte Carlo simulations yield - for the 
$XY$ model - the BKT critical temperature 
$T_{BKT}=0.898J/k_B$ \cite{gupta88}.
When $U \ll J$, 
a BKT transition still occurs at a slightly 
lower critical temperature $T_{BKT}(U)$. 
To evaluate more accurately the effects of the interaction 
term on the BKT critical temperature one may use a
semiclassical approximation to evaluate a renormalized Josephson energy
\cite{jose84,rojas96}: 
the renormalization group equations yields then  
the following equation for ${\cal K} \equiv \beta T_{BKT}(U)$
\begin{equation}
2-\pi {\cal K}  \Bigr[1-\frac{1}{4 {\cal K}}-\frac{1}{X_u 
{\cal K}^3} \, g({\cal K},X_u) \Bigr]=0
\label{radice}
\end{equation}
where $X_u=U/\pi J$ and $g({\cal K},X_u) \equiv \sum_{n=1}^{\infty} 
[X_u {\cal K}^2 / 2-(n^2/2) 
ln (1+ X_u {\cal K}^2/n^2)]$. One may see \cite{smerzi04} that 
for $U/J<36/\pi$, Eq.(\ref{radice}) has an {\it unique} solution; 
at variance, for $U/J>36/\pi$ Eq.(\ref{radice}) 
does not have any solution. 
The value $U/J=36/\pi$ is in reasonable agreement with the known 
mean-field prediction $2 z J / U \approx 1$ 
for the $T=0$ phase transition 
(see, e.g., Eq. (30) of Ref. \cite{vanoosten01} for $N_0 \gg 1$). 
In the inset of Fig.1 we plot the value of $T_{BKT}(U)/T_{BKT}$ 
(where $T_{BKT}$ is the BKT critical temperature of the XY model 
with $U=0$) as a function of $U/J$ from the numerical solution 
of Eq.(\ref{radice}). 

The emerging physical picture is the following: 
There are two relevant temperatures for the system, 
the temperature $T_{BEC}$, at which in each well there is a condensate, 
and the temperature $T_{BKT}$ at which the condensates phases 
start to coherently point in the same direction. Of course, 
in order to have well defined condensates phases one should have 
$T_{BKT} < T_{BEC}$.
The critical temperature $T_{BEC} \approx 0.94 \hbar N_0^{1/3} 
\bar{\omega}/k_B$ where $\bar{\omega}= (\tilde{\omega}_r^2 \omega_z)^{1/3}$.
With the numerical values given in \cite{greiner01}, 
$T_{BEC}$ turns out to be  
$ \gtrsim \ 500 nK$ for $s \gtrsim 20$. When $T < T_{BEC}$, 
the atoms in the well $\vec{j}$ 
of the $2D$ optical lattice may be described by a macroscopic 
wavefunction $\psi_{\vec{j}}$. 
Furthermore, when the fluctuations around 
the average number of atoms per site $N_0 \gg 1$ are strongly suppressed, 
one may put, apart from the factor $\sqrt{N_0}$ constant across the array,  
$\psi_{\vec{j}}=e^{i \theta_{\vec{j}}}$. 
The temperature $T_{BKT}$ is of order 
of $J/k_B$: with the experimental parameters of 
\cite{greiner01} and $V_0$ between $20$ and $25E_R$, 
one has that $T_{BKT}$ is between $10$ and $30nK$, which is 
sensibly smaller than the condensation temperature 
$T_{BEC}$ of the single well. 
A similar picture describes  
also $2D$ arrays of superconducting Josephson junctions 
\cite{simanek94,fazio01}: they 
exhibit a temperature $T_{BCS}$ at which the metallic grains 
placed on the sites of the array become (separately) superconducting 
and the Cooper pairs may be described by macroscopic wavefunctions. 
At a temperature $T_{BKT}<T_{BCS}$, 
the array undergo a BKT transition and the system - as a whole - 
becomes superconducting.

The experimental signature of the BKT transition in bosonic planar lattices 
is obtained by measuring, as a function of the temperature, 
the central peak of the interference pattern 
obtained after turning off the confining potentials \cite{cm}. 
In fact, the peak of the momentum distribution at 
$\vec{k}=0$ is the direct analog of the magnetization of a 
finite size $2D$ $XY$ magnet. For the $XY$ magnets, the spins can be written 
as $\vec{S}_{\vec{j}}=(\cos{\theta_{\vec{j}}},\sin{\theta_{\vec{j}}})$ and 
the magnetization is defined as $M=(1/N) \cdot  
\langle \, \mid \sum_{\vec{j}} \vec{S}_{\vec{j}} \mid \, \rangle$ 
where $\langle \cdots \rangle$ stands for the thermal average. 
A spin-wave analysis at low temperatures yields 
$M=(2N)^{-k_B T/8 \pi J}$ \cite{bramwell93,berezinskii73}. 
With discrete BECs at $T=0$, all the phases 
$\theta_{\vec{j}}$ are equal at the equilibrium and 
the lattice Fourier transform of $\psi_j$, 
$\tilde{\psi}_{\vec{k}}=\frac{1}{N} 
\sum_{\vec{j}} \psi_{\vec{j}} \, \, \, e^{-i \vec{k} \cdot {\vec{j}}}$, 
exhibits a peak at $\vec{k}=0$ ($\vec{k}$ 
is in the first Brillouin zone of the $2D$ square lattice) and
the magnetization is: 
\begin{equation}
M=\langle \, \mid \tilde{\psi}_0 \mid \, \rangle.
\label{M}
\end{equation}  

The intuitive picture of the BKT transition is then the following: 
at $T=0$, all the spins point in the same direction.  
Increasing the temperature, bound vortex pairs are thermally induced.  
In Fig.2, left inset, 
a single free vortex is depicted: we plot a configuration 
of the phases such that $\theta_{\vec{j}}$ equals the polar angle of the 
site $\vec{j}$ with respect to the core vortex 
(notice that with $\theta_{\vec{j}}$ equals 
the polar angle plus $\pi / 2 $ one would have 
the usual plot of a vortex). 
As one can see, a single vortex modifies the phase 
distribution also much far from its core and the square modulus of 
its lattice Fourier transform $\mid \tilde{\psi}_{\vec{k}} \mid^2$
has a minimum at $\vec{k}=0$. At variance, a vortex-antivortex  
pair [see Fig.2, right inset] 
modifies the phase distribution only near the center of 
the pair (in this sense is a {\em defect}) and its lattice Fourier transform 
has a maximum at $\vec{k}=0$. 
Increasing the temperature, vortices are thermally induced. 
For $T < T_{BKT}$ only bound vortex pairs are present, and in
average the spins continue to point in the same direction. 
When the condensates expand, a large peak (i.e., a magnetization) 
is observed in the central $\vec{k}=0$ momentum component. 
Rising further the temperature, 
due to the increasing number of vortex pairs, the central peak density 
decreases from the $T=0$ value. For $T \approx T_{BKT}$ 
the pairs starts to unbind and free vortices begin to appear, determining 
a sharp decrease around $T_{BKT}$ of the magnetization. 
At high temperatures, only free vortices are present,
leading to a randomization 
of the phases and to a vanishing magnetization. 
In Fig.1 we plot the intensity of the central peak of the 
momentum distribution (normalized to the value at $T=0$) in a 2D 
lattice as a function of the reduced temperature $k_B T/J$, 
evidencing the sharp decrease 
of the magnetization around the critical temperature \cite{cm}. 

 
At $t=0$ the amplitude of the $\vec{k}=0$ peak of the 
momentum distribution is simply given by the thermal average 
of $\tilde{\psi}_{0}$. By measuring the $\vec{k}=0$ 
peak (i.e., $\langle \, \mid \tilde{\psi}_0  \mid^2 \,    \rangle$) 
at different temperatures, one obtains the results plotted in Fig.1. 
The figure has been obtained using a Monte Carlo simulation of the 
$XY$ magnet for a $40 \times 40$ array: we find $k_B T_C \approx 1.07 J$. 
In Fig.1 we also plot the low-temperature 
spin wave prediction \cite{berezinskii73} (solid line). 
At times different from  
$t=0$, the density profiles 
are well reproduced by the free expansion of the ideal gas. 
One obtains $\tilde{\psi}(\vec{p},t)=\chi(p_z) 
e^{-i[(p_z+mgt)^3-p_z^3]/6m^2g\hbar} \tilde{\varphi}(p_x,p_y,t)$ 
[where $\chi(p_x) \propto  
e^{-\sigma_x^2 p_x^2/ 2 \hbar^2}$ and similarly for $\chi(p_y)$ and  
$\chi(p_z)$], i.e. 
a uniformly accelerating motion along $z$ and a free motion in the 
plan $x-y$, with $\tilde{\varphi}(p_x,p_y,t)= 
\chi(p_x) \chi(p_y) \tilde{\psi}_{\vec{k}}e^{-i (p_x^2+p_y^2) t/2\hbar m}$
giving the central and lateral peaks of 
the momentum distribution as a function of time for different 
temperatures. 

An intense experimental work is now focusing on the 
Bose-Einstein condensation in two dimensions: 
at present a crossover to two-dimensional behaviour 
has been observed for $Na$ \cite{gorlitz01} and $Cs$ atoms \cite{hammes03}. 
Our analysis relies on two basic assumptions; namely, 
the validity of the tight-binding approximation 
for the Bose-Hubbard Hamiltonian and the requirement 
that the condensate in the optical lattice may be regarded as planar 
\cite{petrov00}. It is easy to see that the first assumption 
is satisfied if, at $T=0$, $V_0 \gg \mu$ 
(where $\mu$ is the chemical potential), and, at finite temperature, 
$\hbar \tilde{\omega}_r \gtrsim k_B T$. The second assumption is much more 
restrictive since it requires freezing the transverse excitations; 
for this to happen one should require a condition on the transverse 
trapping frequency $\omega_z$. Namely, one should have that, at $T=0$, 
$\hbar \omega_z \gtrsim 8 K$ and that, at finite temperature, 
$\hbar \omega_z \gtrsim k_B T$ (since $\omega_z \ll \tilde{\omega}_r$, 
the latter condition also implies that $\hbar \tilde{\omega}_r 
\gtrsim k_B T$). In \cite{greiner01} it is  
$V_0=s \cdot k_B \cdot 0.15 \mu K$ and for $s \gtrsim 20$ 
the tight-binding conditions are satisfied since 
$10 Hz \gtrsim 8K/2\pi \hbar$; furthermore, if $\omega_z=2 \pi \cdot 1kHz$, 
one may safely regard our finite temperature analysis to be valid at least 
up to $T \sim 50 nK$. We notice that the experimental signature for the  
BKT transition for a continuous 
(i.e., without optical lattice) weakly
interacting 2D Bose gas \cite{prokofev01} is also given 
by the central peak of the atomic density of the 
expanding condensates.

Our study evidences the possibility that Bose-Einstein
condensates loaded on a $2D$ optical lattice may exhibit 
- at finite temperature - a new coherent 
behaviour in which all the phases of the condensates located in each well 
of the lattice point in
the same direction. The finite temperature transition, which is due 
to the thermal atoms in each well, is mediated by vortex defects and may 
be experimentally detectable by looking at the
interference patterns of the expanding condensates. Our analysis
strengthens - and extends at finite temperature - the 
striking and deep analogy of bosonic planar systems 
with $2D$ Josephson junction arrays \cite{anderson98}. 


{\em Acknowledgements} 
This work has been supported by MIUR through grant No. 2001028294 
and by the DOE.

\begin{figure}
\centerline{\psfig{figure=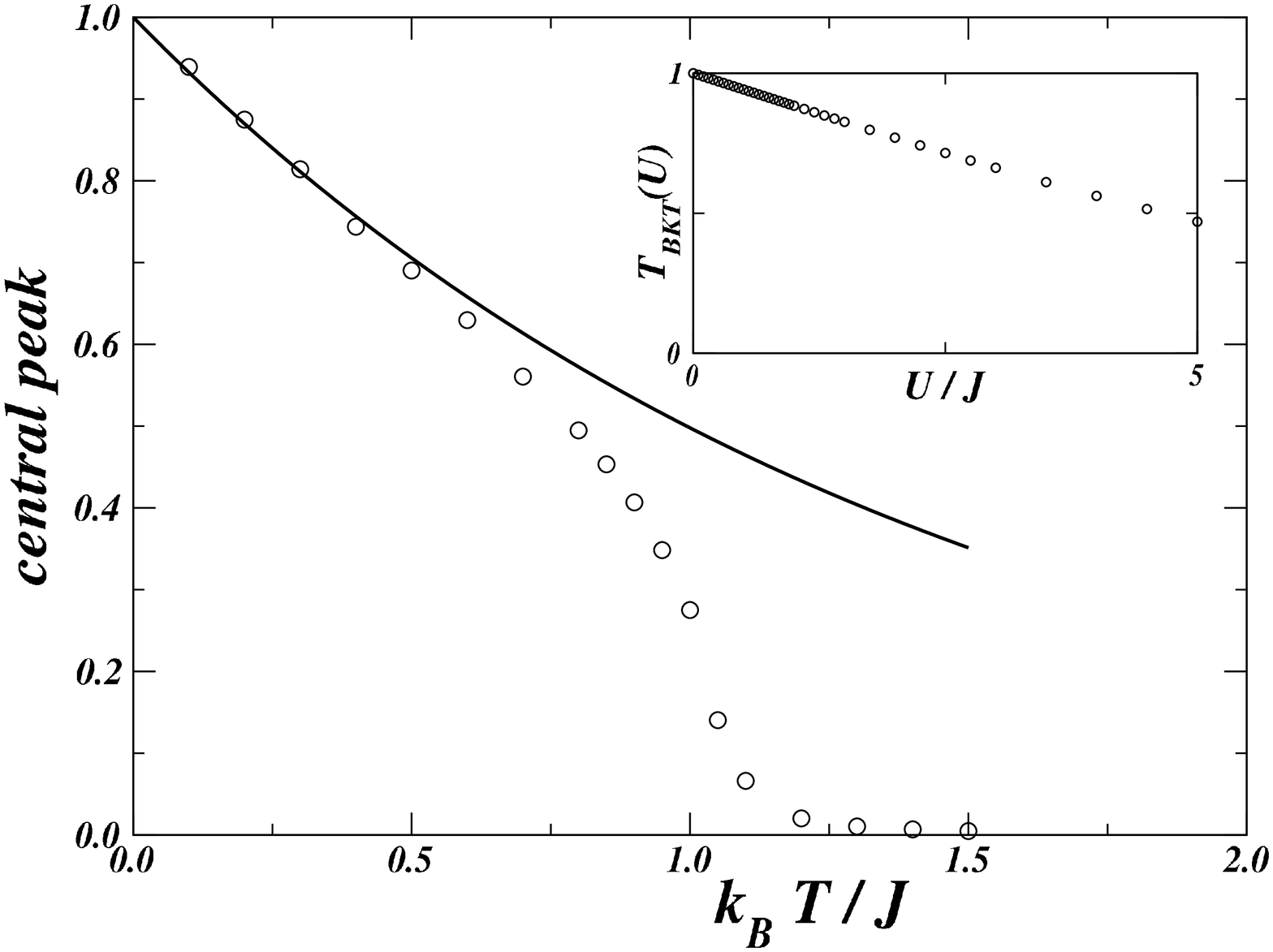,width=60mm,angle=270}}
\caption{Intensity of the central peak of the momentum distribution 
(normalized to the value at $T=0$) as a function of the reduced temperature 
$t=k_B T/J$; empty circles: Monte Carlo simulations; 
solid line: low-temperature spin wave prediction 
\protect\cite{berezinskii73}. 
Inset: the BKT critical temperature $T_{BKT}(U)$ 
(in units of the $XY$ critical temperature) as a function  
of $U/J$ from Eq.(\ref{radice}).}
\label{fig1}
\end{figure}

\begin{figure}
\centerline{\psfig{figure=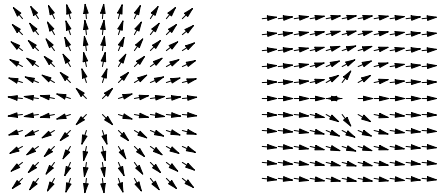,width=60mm,angle=0}}
\caption{Left: a lattice vortex; right: a vortex-antivortex 
pair.}
\label{fig2}
\end{figure}


\end{document}